\documentclass[aps,twocolumn,prx,preprintnumbers,showpacs,amsmath,amssymb,superscriptaddress]{revtex4-1}
\usepackage[dvipdfm, colorlinks, citecolor=blue]{hyperref}
\usepackage{graphicx}
\usepackage{dcolumn}
\usepackage{threeparttable}
\usepackage{multirow}
\usepackage{booktabs}
\usepackage{txfonts}
\usepackage{xcolor}
\usepackage{bm}
\usepackage{amssymb}
\usepackage{amsmath}
\usepackage{latexsym}
\usepackage{epsfig}
\usepackage{amsbsy}
\usepackage{array}
\usepackage{tabularx}

\newcommand{\ket}[1]{\mid #1 \rangle }

\begin{document}

\title{Electron and hole doping in the relativistic Mott insulator Sr$_2$IrO$_4$: \\
a first-principles study using band unfolding technique}

\author{Peitao Liu}
\affiliation{University of Vienna, Faculty of Physics and Center for Computational Materials Science, Sensengasse 8/8,
A-1090 Vienna, Austria}
\affiliation{Shenyang National Laboratory for Materials Science, Institute of Metal Research,
Chinese Academy of Sciences, Shenyang 110016, China}

\author{Michele Reticcioli}
\affiliation{University of Vienna, Faculty of Physics and Center for Computational Materials Science, Sensengasse 8/8,
A-1090 Vienna, Austria}

\author{Bongjae Kim}
\affiliation{University of Vienna, Faculty of Physics and Center for Computational Materials Science, Sensengasse 8/8,
A-1090 Vienna, Austria}

\author{Alessandra Continenza}
\affiliation{Department of Information Engineering, Computer Science and Mathematics, University of L'Aquila, Via Vetoio, 67100 Coppito (AQ), Italy}

\author{Georg Kresse}
\affiliation{University of Vienna, Faculty of Physics and Center for Computational Materials Science, Sensengasse 8/12,
A-1090 Vienna, Austria}

\author{D.D. Sarma}
\affiliation{Solid State and Structural Chemistry Unit, Indian Institute of Science, Bangalore-560012, India}

\author{Xing-Qiu Chen}
\affiliation{Shenyang National Laboratory for Materials Science, Institute of Metal Research, Chinese Academy of Sciences, Shenyang 110016, China}

\author{Cesare Franchini}
\email[Corresponding author: ]{cesare.franchini@univie.ac.at}
\affiliation{University of Vienna, Faculty of Physics and Center for Computational Materials Science, Sensengasse 8/8,
A-1090 Vienna, Austria}

\begin{abstract}

We study the effects of dilute La and Rh substitutional doping on the electronic structure of the relativistic Mott insulator
Sr$_2$IrO$_4$ using fully relativistic and magnetically non-collinear density functional theory with the inclusion of an on-site Hubbard $U$ (DFT+$U$+SOC).
To model doping effects, we have adopted the supercell approach, that allows for a realistic treatment of structural relaxations
and electronic effects beyond a purely rigid band approach. By means of the band unfolding technique we have computed the spectral function
and constructed the effective band structure and Fermi surface (FS) in the primitive cell, which are readily comparable with available
experimental data. Our calculations clearly indicate that La and Rh doping can be interpreted as effective electron and (fractional) hole
doping, respectively. We found that both electron and hole doping induce an insulating-to-metal transition (IMT)
but with different characteristics.
In Sr$_{2-x}$La$_x$IrO$_4$ the IMT is accompanied by a moderate renormalization
of the electronic correlation substantiated by a reduction of the effective on-site Coulomb repulsion $U-J$ from 1.6 eV ($x=0$) to
1.4 eV (metallic regime $x=12.5\%$). The progressive closing of the relativistic Mott gap leads to the emergence of connected elliptical
electron pockets at  ($\pi$/2,$\pi$/2) and less intense features at $X$ in the Fermi surface.
The average ordered magnetic moment is slightly reduced
upon doping but the canted antiferromagnetic state is perturbed in the Ir-O planes located near the La atoms.
The substitution of Ir with the nominally isovalent Rh is accompanied by a substantial hole transfer from the Rh site to the nearest neighbor
Ir sites. This shifts down the chemical potential, creates  almost circular disconnected hole pockets in the FS and establishes the emergence
of a two-dimensional metallic state formed by conducting Rh-planes intercalated by insulating Ir-planes.
Finally, our data indicate that hole doping causes a flipping of the in-plane net ferromagnetic moment in the Rh plane
and induces a magnetic transition from the AF-I to the AF-II ordering.

\end{abstract}

\maketitle

\section{INTRODUCTION}

The recently reported spin-orbital $J_\text{eff}=1/2$ Mott insulating state in Sr$_2$IrO$_4$~\cite{Kim2008, Kim2009}
has stimulated a lot of fundamental research aiming to clarify the underlying mechanism responsible for this novel state of matter
which arises from the cooperative interplay of the crystal field, spin-orbit coupling (SOC),  electron-electron
interaction ($U$), and different types of simultaneously active spin-exchange interactions~\cite{Jackeli2009, Khaliullin2012, Kim2014exciton, Liu2015, Hou2016}.
An important aspect of Sr$_2$IrO$_4$ that also has attracted considerable attention is its similarity with high-$T_\text{c}$ cuprate superconductors such
as La$_2$CuO$_4$: these two compounds share the same quasi-two-dimensional layered perovskite structure, Ir and Cu have a nominal $d^5$ and $d^9$
configuration with one effective hole per site and both compounds are described by a $S(J_\text{eff})=1/2$ antiferromagnetic (AF) Heisenberg
model~\cite{cuprate-1, FujiyamaPRL2012, Wang2011, Watanabe2013}. Theoretical studies reporting the possible onset of
superconductivity in electron-doped Sr$_2$IrO$_4$~\cite{Wang2011,Watanabe2013} have spurred an immediate experimental research
that have reported the existence of unusual (pseudogap) metallic states in  Sr$_{2-x}$La$_x$IrO$_4$,
somehow similar to  high-$T_c$ cuprates, but no sign of superconductivity has been found down to very low temperatures for supposedly optimal
doping~\cite{KimScience2014,Yan2015,Kim_n2015,La_Torre2015}.
On whether the Fermi surface (FS) is formed by disconnected Fermi arcs or Fermi lenses is still experimentally uncertain.
More generally, electron and hole doping in (relativistic) Mott insulators are one of the most studied and controversial issues in solid state
theory due to the complex impact that excess carriers can have on the competing spin/charge/orbital interactions~\cite{Lee2006}.
In Sr$_2$IrO$_4$ the study of doping effects can provide important insights on the robustness and tunability of the
spin-orbital Mott state and of the canted in-plane antiferromagnetic ordering. This is associated with a controlled doping-induced modulation
of the dominant interactions, in particular SOC and $U$, as well as with possible changes of the atomic positions (structural distortions), that can
have a strong repercussion on the electronic and magnetic properties.

Among the different forms of electron and hole doping tested in this system, the most studied ones are the substitution of Sr$^{2+}$ with  La$^{3+}$ (electron doping)
and the homovalent heterosubstitution of 5$d^5$ Ir$^{4+}$ with 4$d^5$ Rh$^{4+}$ (effective hole doping)~\cite{Klein08, La_Torre2015, La_Ge2011,All_Lee2012,Rh_Qi2012,Rh_Cao2014, Rh_Clancy2014,
Rh_Ye2015,La_Chen2015,LaRh2015,Ru_Calder2015}. In one of the first studies, Lee and coworkers conducted a systematic investigation of La, Rh, and Ru doping
in Sr$_2$IrO$_4$ by means of optical spectroscopy and found that the $J_\text{eff}=1/2$ state remains spectroscopically robust upon doping and that in
all cases doping induces an insulator-to-metal transition (IMT) for moderately low dopant concentrations ($\approx$ 5\%)~\cite{All_Lee2012}.
They have also suggested that the IMT is associated with subtle alterations of the strength of $U$ and SOC~\cite{All_Lee2012}. This hypothesis has been
studied more recently in a combined angle-resolved photoelectron spectroscopy (ARPES) and tight-binding study of La-doped Sr$_2$IrO$_4$
where the authors have concluded that an optimal agreement between experiment and calculation can be achieved assuming a fast
quenching of $U$ from 2~eV to 0~eV across the transition~\cite{La_Torre2015}. On the other hand, the idea that a reduction of the SOC should be responsible for
the closing of the Mott gap in the Rh-doped sample has been revised by Cao and coworkers: they have shown that  Rh doping  effectively corresponds
to hole doping and therefore also in this case the IMT can be explained by a  band-filling mechanism~\cite{Rh_Cao2014}.
In our study we reconcile these two apparently conflicting interpretations and will show that the effective hole doping mechanism is assisted by the different
SOC strength in Rh and Ir.

\begin{figure}[t!]
\begin{center}
\includegraphics[width=0.48\textwidth]{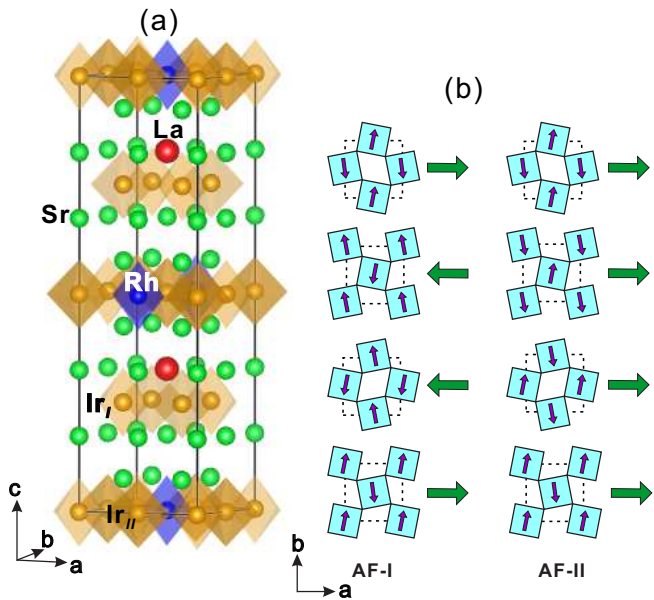}
\end{center}
\caption{(color online) (a) Ball and stick model of supercell adopted in the present study to model $\rm Sr_{2-x}La_xIr_{1-y}Rh_yO_4$,
showing the substituional Sr$\rightarrow$La (dark spheres) and Ir$\rightarrow$Rh (dark octahedra) sites. Two different kinds of Ir
atoms can be identified: Ir$_{II}$, nearest neighbor to Rh, and Ir$_{I}$. Oxygen atoms are not displayed.
(b) Schematic plot of the canted AF-I and AF-II orderings, showing both the ordered AF moments (thin blue arrows)
and the flipping of the net in-plane ferromagnetic moment (thick green arrows).}
\label{fig:mag}
\end{figure}

Substitutional doping has also a significant influence on the magnetic order. Upon La doping, a complex phase diagram was proposed
showing the weakening of the long-range antiferromagnetic (AF) order and the appearance of spin-glass-like  phase beyond a critical
La concentration~\cite{La_Chen2015}. Rh doping was found to induce a magnetic transition from the AF-I to the AF-II state
involving the in-plane flipping of the net ferromagnetic (FM) moment~\cite{Rh_Clancy2014,Rh_Ye2015} as shown in  Fig.~\ref{fig:mag}(b).

Despite the relatively large amount of experimental studies on doped Sr$_2$IrO$_4$ first-principles calculations were
not reported to date. In this work, we study the evolution of the spin canted Mott state in Sr$_2$IrO$_4$ as a function of
La and Rh doping using relativistic density functional theory with the inclusion of an on-site Hubbard $U$, the same scheme that we
have used for the description of the ground state electronic and magnetic interactions in the undoped compound~\cite{Liu2015}.
Doped Mott insulators are generally studied using model Hamiltonian schemes as these methods are designed to describe
the role of strong correlation across the IMT, in particular band renormalization and the transfer of spectral weight, which are not
accessible by single-particle approaches. However, the combined ARPES and  tight-binding study  of La Torre {\em et al.}~\cite{La_Torre2015}
indicates that the band structure of doped Sr$_2$IrO$_4$ is sufficiently well described using single-particle-like dispersions suggesting
that a single-particle band theory like DFT+$U$ might be a suitable approximation in this case. We
will carefully address this point in this work.

From an {\em {ab initio}} perspective the {\em{realistic}} treatment of doping requires the employment of large supercells (SC)
and for a one-to-one comparison with available ARPES electronic structure it is necessary to map the band structure calculated
in the supercell into an effective band structure (EBS) projected in the Brillouin zone of the primitive cell. To this aim we used the band structure unfolding method
introduced few years ago by Popescu and Zunger~\cite{Popescu2010,Popescu2012} and recently incorporated in the Vienna \emph{Ab initio} Simulation
Package (VASP)~\cite{Eckhardt2014, Michele2015}. With this approach we have computed the EBS and FS of La- and Rh-doped
Sr$_2$IrO$_4$ at different concentrations across the IMT and the obtained results are in reasonable agreement with the available ARPES spectra.
Our data suggest that La doping leads to a moderate reduction of the effective Coulomb interaction $U_\text{eff}$=$U-J$ from
1.6 eV, optimum value in the undoped system, to 1.4~eV at the electron doping concentration $x=12.5\%$. The upshift of the chemical potential upon electron doping and
the renormalization of the electronic correlation result in a metallic state, however, we
fail to predict the merging of the lower and upper Hubbard band at low electron doping concentrations at the optmium $U_\text{eff}$ values.
The complete closing of the gap is achieved for smaller values of $U_\text{eff}$ ($\approx$ 0.8~eV).
As already mentioned, although Rh in Sr$_2$RhO$_4$ is isoelectric with Ir in Sr$_2$IrO$_4$, our DFT results confirm the experimental conclusions that Rh
substitution in Sr$_2$IrO$_4$  acts as a hole donor shifting up the valence bands and creating a hole pocket around the $X$ point.

In the next section we will provide the technical details on the computational methods and setup. The results will be presented and discussed in
Sec.~\ref{sec_res} and summarized in Sec.~\ref{sec:sum}.

\begin{figure*}[t!]
\begin{center}
\includegraphics[width=0.80\textwidth]{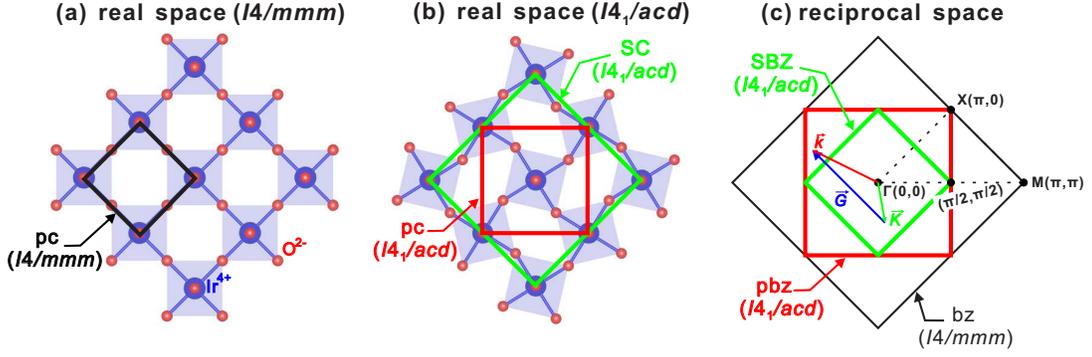}
\end{center}
\caption{(color online) The relation between primitive cell (pc) and supercell (SC)
in real space for (a) undistorted ($I4/mmm$) and (b) distorted ($I4_1/acd$) structure.
(c) The corresponding primitive (pbz) and supercell (SBZ) Brillouin zones,
and their associated wave vectors $\vec{k}$ and $\vec{K}$ connected by the reciprocal lattice
vector $\vec{G}$ of the supercell,
are illustrated. Note that in (c) the fractional coordinates of high-symmetry
points are displayed using $I4/mmm$ notation in units of 1/$a_0$ with $a_0$ being
the $ab$-plane lattice constant of the undistorted structure. }
\label{str_BZ}
\end{figure*}

\section{Computational details}

Our first-principles calculations including $U$ and SOC were performed using the
projector augmented wave method~\cite{PAW} as implemented in VASP~\cite{Kresse-1, Kresse-2}.
The Perdew-Burke-Ernzerhof (PBE)~\cite{PBE} approximation was employed for the
exchange-correlation functional. Plane waves have been included
up to an energy cut off of 400 eV. 4$\times$4$\times$2 $\vec{k}$-point grids
were used to  sample the Brillouin zone.
For the undoped case, we used an effective $U_\text{eff}$ = $U-J$ = 1.6 eV on the Ir site,
computed using the constrained random phase approximation (cRPA), which yields a very good account
of the ground state electronic and magnetic properties~\cite{Liu2015}. The cRPA has been also employed to compute
the $U_\text{eff}$ in the high-doping regime ($x=12.5~\%$).
For the Rh doping case, we have included an on-site $U_\text{eff}$=1.2~eV on the Rh sites,
estimated by cRPA~\cite{Martins2011}.

In order to fully account for the doping effects on the electronic structure
of doped Sr$_2$IrO$_4$, we employed the supercell approach.
The supercell ($\sqrt 2 \times \sqrt 2  \times 1$) with 112 ions, depicted in  Fig.~\ref{fig:mag}(a),
was constructed based on the experimental unit cell with $I4_1/acd$ symmetry~\cite{Crawford1994}, as shown in Fig.~\ref{str_BZ}(b).
For the La doping (La$^{3+}$ for Sr$^{2+}$)~\cite{La_Ge2011}
the chemical formula becomes Sr$_{32-y}$La$_y$Ir$_{16}$O$_{64}$
with $y$ being the number of  La$^{3+}$ ions in the supercell. Thus, substitution of one (two) Sr$^{2+}$ by La$^{3+}$
corresponds to a La doping concentration $y$ of 3.125\% (6.25\%). It should be noted that
the nominal electron doping counted per Ir site is doubled due to the stoichiometry of Sr$_2$IrO$_4$~\cite{La_Torre2015},
yielding electron doping concentrations of  $x$=2$y$=6.25\% and 12.5\%, respectively. With respect to the concentration $x$,
the general chemical formula for La doping reads $\rm Sr_{2-x}La_xIrO_4$.
For the Rh doping (Rh$^{3+}$ for Ir$^{4+}$ in the supercell $\rm Sr_{32}Ir_{16-x}Rh_xO_{64}$)~\cite{Rh_Cao2014}, substitution of one (two)
Ir$^{4+}$ by Rh$^{3+}$ results in nominal hole doping on the Ir site of $x$=6.25\% (12.5\%).
To avoid confusion, we use $x$ to label the electron (hole) doping concentration throughout the paper.
Unless explicitly stated, all supercell calculations
were performed by fully relaxing the atomic positions at fixed volume, corresponding to the experimental
volume of the undoped compound, in order to preserve local structural effects of the dopant atoms.

\begin{figure*}[t!]
\begin{center}
\includegraphics[width=1.0\textwidth]{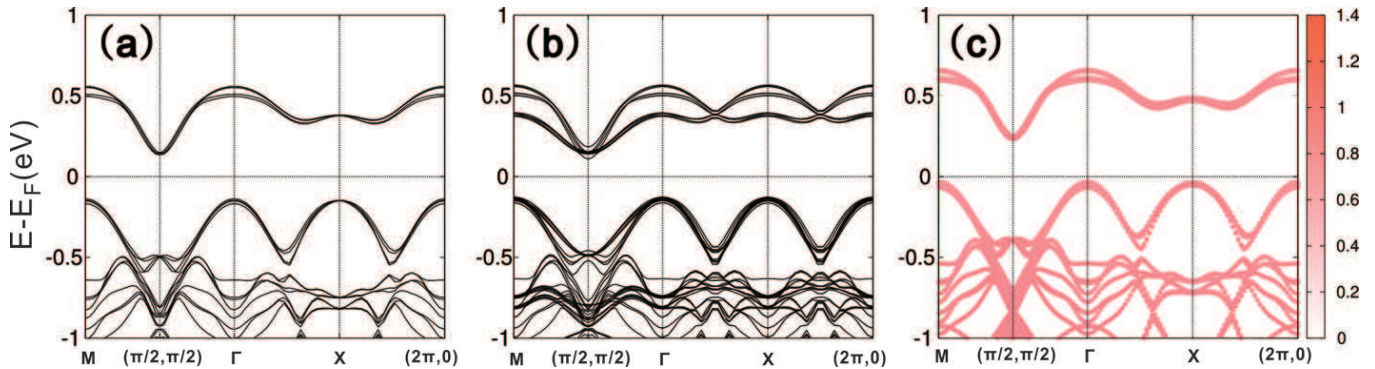}
\end{center}
\caption{(color online) Band structure of (a) primitive cell and (b) supercell.
(c) EBS calculated from the unfolding method. Note that the color bar in (c) represents
the Bloch character given by Eq. (\ref{P_G}). The high-symmetry points here are
consistent with the primitive $I4/mmm$ notation displayed in Fig.~\ref{str_BZ}(c).}
\label{SC_band}
\end{figure*}

To analyze the effects of doping on the band structure of the employed supercells,
we have adopted the unfolding technique recently implemented in VASP~\cite{Eckhardt2014, Michele2015}.
In fact, calculations based on the supercell approach, with a unit cell $N$~times larger than the primitive cell, lead to a down folded Brillouin zone that, in most cases,
makes it difficult to  interpret directly the resulting band structures.
An unfolding technique is required in order to obtain a clearer description of the band structure.
The relation
\begin{equation}
\vec{k}+\vec{g}=\vec{K}+\vec{G}
 \label{eq_fold}
\end{equation}
describes the folding of the reciprocal space, mapping a wave vector $\vec{K}$ of the supercell
 into $N$ wave vectors $\vec{k}$ in the Brillouin zone of the primitive cell (pbz), by means of the reciprocal lattice vectors
 $\vec{g}$ and $\vec{G}$ of the primitive and supercell, respectively [see Fig.~\ref{str_BZ}(c)].
The projection $P_{\vec{K}m}(\vec{k})$ of the eigenstates $\ket{\Psi_{\vec{K}m}}$ of the supercell into
eigenstates $\ket{\psi_{\vec{k}n}}$ of the primitive cell, where $m$ and $n$ are energy band indexes,
provides an effective band structure (EBS) in the pbz starting from eigenvalues calculated in the Brillouin zone of the  supercell (SBZ).
As proved by Popescu and  Zunger~\cite{Popescu2010,Popescu2012}, the projection $P_{\vec{K}m}(\vec{k})$,
called Bloch character, can be written in terms of the plane wave coefficients of the supercell eigenstates only.
In fact, given Eq.~(\ref{eq_fold}), the eigenstates of the supercell can be written as
\begin{equation}
 \begin{split}
  \Psi_{\vec{K}m}(\vec{r}) &= \sum_{\vec{G}} C_{\vec{K}+\vec{G},m} e^{i(\vec{K}+\vec{G}) \cdot \vec{r}} \\
  &= \sum_{\vec{k}} \sum_{\vec{g}} C_{\vec{k}+\vec{g},m} e^{i(\vec{k}+\vec{g}) \cdot \vec{r}}~.
 \end{split}
\end{equation}
Therefore, the Bloch character can be obtained from the supercell calculation alone and no primitive cell calculations are required, because of the relation:
\begin{equation}
 \begin{split}
  P_{\vec{K}m}(\vec{k})
  &=\sum_n|\langle\Psi_{\vec{K}m}|\psi_{\vec{k}n}\rangle|^2 \\
  &=\sum_{\vec{g}}   |C_{\vec{k}+\vec{g}, m}|^2~.
  \label{P_G}
 \end{split}
\end{equation}
The spectral function $A(\vec{k},E)$ can be hence calculated as
\begin{equation}
A(\vec{k},E)=\sum_m P_{\vec{K}m}(\vec{k})\delta(E_m-E)~,
\label{eq2_A}
\end{equation}
providing a directly comparison of the calculated EBS with ARPES experiments in the reciprocal space of the primitive cell.

To validate and demonstrate the applicability of the unfolding scheme, we show in Fig.~\ref{SC_band} a comparison between the
ground state band structure of undoped Sr$_2$IrO$_4$ in the primitive $I4_1/acd$ cell [Fig.~\ref{SC_band}(a)] and in the
$I4_1/acd$ supercell [Fig.~\ref{SC_band}(b)], together with the EBS of the supercell unfolded into the primitive cell [Fig.~\ref{SC_band}(c)].
The primitive cell band structure exhibits the well known relativistic Mott $J_\text{eff} = 1/2$ state formed by one filled and one empty band,
usually referred to as lower and upper relativistic Mott-Hubbard bands (LHB and UHB, respectively),
which are separated by a small band gap of about 0.3~eV~\cite{Kim2008}.
The EBS of the undoped supercell shown in Fig.~\ref{SC_band}(c) reproduces well this behavior: the EBS are formed by
sharp bands reflecting the fact that the eigenstates of the primitive cell can contribute either fully or not at all
to a given supercell eigenstate.
Dopant atoms will clearly disrupt this one-to-one correspondence between supercell and primitive cell eigenstates,
introducing mixed contributions represented by intermediate values for the Bloch characters and, hence, broadened bands.

\section{Results and Discussion}\label{sec_res}

\subsection{La doping}

We start this section by discussing the relation between doping and electron-electron correlation.
The ARPES spectra clearly indicate a fast collapse of the relativistic Mott gap and the appearance of a large Fermi surface
at relatively low doping levels~\cite{La_Torre2015}.
As La doping effectively induces excess electrons in the system, one would expect
that the strength of the electron-electron correlation measured in terms of $U_\text{eff}$ should decrease with
increasing doping concentration as a consequence of the increased screening from the metallic states.
Considering that the IMT occurs at rather low doping (few \%) one would expect that the $U_\text{eff}$ would not change much across the transition.
In contrast to this expectation, however, previous tight-binding calculations suggested that only a complete quenching of the Coulomb
repulsion $U$ from an ideal value of 2.0 eV, assumed for the undoped case, to 0 eV at $x=0.1$ reproduced the experimental ARPES data.
To verify the reliability of this conjecture we have computed  $U_\text{eff}$ fully \emph{ab intio} using cRPA  for the
largest doping concentration considered in our study, $x=12.5~\%$, using the conventional cell containing 56 atoms.
We found  $U_\text{eff}=1.4$~eV implying that $U_\text{eff}$ is only moderately affected by doping at these low
concentrations. Assuming a linear decrease of $U_\text{eff}$ upon doping a full suppression of $U_\text{eff}$ would occurs
at very large doping, $x \approx 80~\%$.
A conceptually similar reduction of electronic correlation effects due to electron
doping, analyzed in terms of the changes of the mass enhancement factor, was recently reported for La-doped Sr$_2$RhO$_4$~\cite{Ahn2015}.

\begin{figure}[h!]
\begin{center}
\includegraphics[clip = , width=0.47\textwidth]{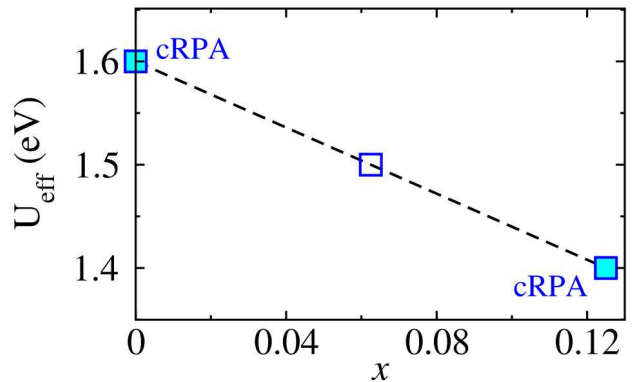}
\end{center}
\caption{(color online) Effective electron-electron interaction $U_\text{eff}$  as a function of electron doping concentrations $x$.
The filled symbols correspond to the actual values computed by cRPA whereas the the open symbol is the interpolated value of
$U_\text{eff}$ for the lowest doping level, $x=6.25\%$.}
\label{Uxm}
\end{figure}

\begin{figure}
\begin{center}
\includegraphics[width=0.5\textwidth]{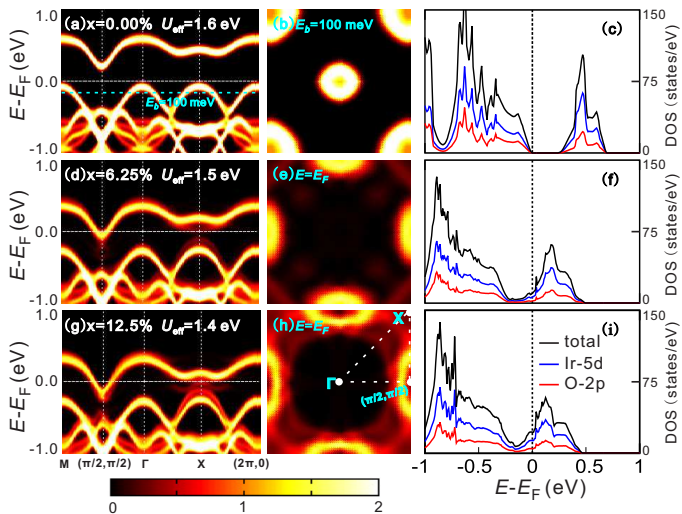}
\end{center}
\caption{(color online) Doping induced IMT in Sr$_2$IrO$_4$ in terms of EBS, FS and total and Ir-5$d$ and O-2$p$
projected DOS at different electron
doping concentration $x$: (a,b,c) $x=0$; (d,e,f) $x$=6.25\%; (g,h,i) $x=12.5\%$.
The Fermi level is set to zero. For the insulating $x=0$ case the FS is replaced by
an isoenergy (E$_b$=100 meV) contour plot.}
\label{band_La}
\end{figure}

\begin{figure*}
\begin{center}
\includegraphics[width=0.95\textwidth]{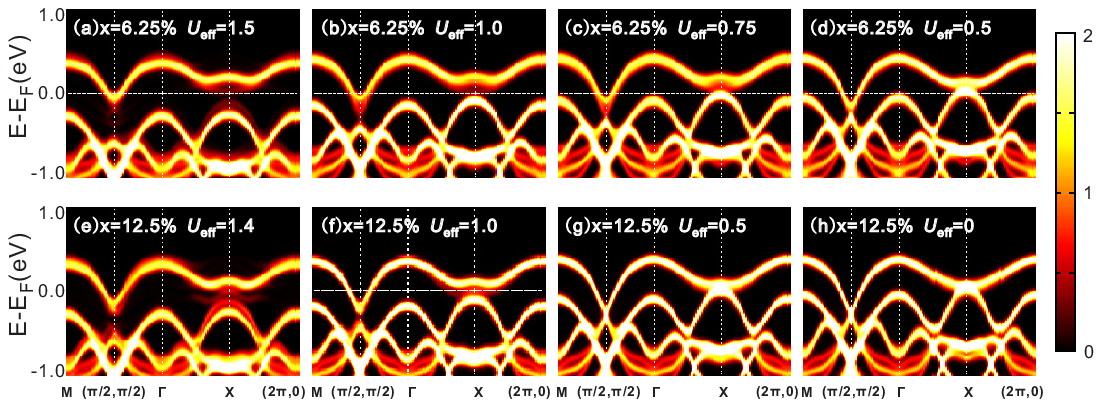}
\end{center}
\caption{(color online) EBS of La-doped Sr$_2$IrO$_4$
as a function of the  effective Hubbard interactions $U_\text{eff}$ (eV) for
the electron doping concentration $x$=6.25\% and 12.5\%.
The color coding indicates the spectral function calculated by Eq. (\ref{eq2_A}).
}
\label{LaU}
\end{figure*}

With this set of \emph{optimum} values of $U_\text{eff}$ at hand, we have computed the
EBS, FS and density of states (DOS) for increasing doping concentrations: $x$=0, 6.25\% and $12.5\%$.
The results, shown in Fig.~\ref{band_La} clearly show that upon La-doping  Sr$_2$IrO$_4$ undergoes
a IMT associated with a progressive reduction of the separation between the LHB and the UHB,
and the emergence of a structured FS.
Already at the lowest doping concentration, $x=6.25\%$, the bottom of the UHB crosses the Fermi energy at
($\pi$/2,$\pi$/2) forming elliptical spots in the FS. By increasing doping the elliptical electron pockets become sharper
and additional features appear in the FS around the $X$ point arising from the doping-induced
spectral weight broadening at the top/bottom of the lower/upper Hubbard bands.
This is in modest agreement with the available ARPES experiments~\cite{KimScience2014,Yan2015,Kim_n2015,La_Torre2015},
specifically some differences are noticeable at the $X$ point and at ($\pi$/2,$\pi$/2).
In the experiments, the distance between the upper and lower Hubbard band closes
upon doping, and as the lower and upper Hubbard band merge, a Dirac cone develops at the ($\pi/2$, $\pi/2$) point.
In our simulations this only starts to happen at much larger concentrations beyond $x$=12.5\% or for smaller values of $U_\text{eff}$
(see Fig.~\ref{LaU}).
Furthermore in the experiments the valence band edge at the $X$ point moves above the Fermi-level.
If we keep the $U_\text{eff}$ largely fixed, this feature is not observed in our theoretical calculations,
though a doping-induced broadening of the lower/upper Hubbard bands is visible.
We believe that the main reason for this erroneous result is the inadequacy of the LDA+$U$ method.
In the present case, the conduction band at the $X$ point is non-localized,
similar to conventional semiconductors. Likewise the top most valence band state at the $X$
point is itinerant. Using a simple local correction $U$ can open the band gap between
the lower and upper Hubbard band, but it is not able to deal properly with the intricate
details of non-localized, itinerant states. The minimum complexity required to describe band gap
narrowing upon doping is $GW$, but $GW$ for the system sizes considered here is yet
very difficult, and we relegated this to future work.

To clarify how $U_\text{eff}$ changes the band structure, we report in  Fig.~\ref{LaU} the EBSs  with projected spectral weight
computed for different values of $U_\text{eff}$  at $x$=6.25\%, and $x$=12.5\%. Specifically, we have considered $U_\text{eff}$=1.5, 1.0, 0.75
and 0.5~eV for $x$=6.25\%, and $U_\text{eff}$= 1.4, 0.8, 0.5, and 0.0~eV for the largest concentration, $x$=12.5\%.
Clearly, the value of $U_\text{eff}$ affects the robustness of the $J_\text{eff}=1/2$ state.
Upon decreasing $U_\text{eff}$ (from left to right in Fig.~\ref{LaU}) the separation between the LHB and the UHB is progressively reduced
as a result of the downward shift of the UHB, especially near the ($\pi$/2,$\pi$/2) point and a upward shift of the
LHB around the $X$ point. Moreover, the reduction of $U_\text{eff}$ strengthens the intensity of the Dirac cone at ($\pi$/2,$\pi$/2) around $-$0.2 eV,
and influences the intensity of the spectral weight at the $X$ and ($\pi$/2,$\pi$/2) points. Obviously reducing
U would indeed reproduce the experimental results, for instance the emergence of the Dirac cone
at the ($\pi$/2,$\pi$/2) and a shift of the valence band edge at $X$ above the Fermi-level.
However, from a first principles perspective this is certainly unsatisfactory, since we
can not justify the rapid reduction of $U$ with doping.

\begin{figure}
\begin{center}
\includegraphics[width=0.48\textwidth]{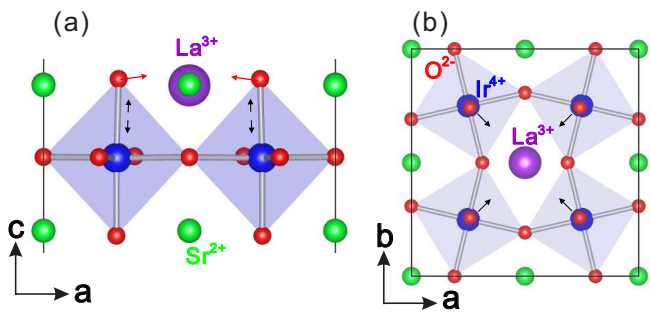}
\end{center}
\caption{(color online) Side (a) and top (b) view of the structural relaxations around the La$^{3+}$ ion manifested
by tilting and stretching of the IrO$_6$ octahedra (highlighted schematically by the arrows).
}
\label{relax_str}
\end{figure}

After discussing the origin of the doping-induced IMT we conclude this section by discussing the effect of La-doping on
the structure and on the magnetic ordering. Due to the smaller ionic radius of La$^{3+}$ compared with Sr$^{2+}$, the inclusion of La$^{3+}$
induces small local distortions ($\approx$ 2\%) near the La site due to the
stronger electrostatic La$^{3+}$-O attraction as compared to Sr$^{2+}$-O leading to a smaller La-O distance as compared to Sr-O in the
undoped sample and to a small expansion of the Ir-O apical bond-length within the IrO$_6$ octahedron,
as schematically shown in Fig.~\ref{relax_str}. The structural modifications of 1-2 unit cells away from the La sites are found to be
almost negligible ($< 1$\%).
We found that the effects of these structural rearrangements on the electronic structure (not shown) are
minimal. Conversely doping and structural relaxation perturbed the ordered canted state characteristic of the undoped phase.
Our data indicate that apart from a relatively small decrease of the average ordered moment,  La-doping leads to a
local disruption of the canted ordering in the Ir planes closer to the La impurity. This is in line with the experimental observation of a
weakening of the long-range canted AF order~\cite{La_Chen2015} upon La-doping.

\subsection{Rh doping}

Following the same procedure adopted for the electron doping case we have studied the evolution of the electronic and magnetic state of
Sr$_2$IrO$_4$ upon hole doping. As mentioned in the introduction effective hole doping can be achieved by replacing $5d$ Ir$^{4+}$ with
the corresponding nominally isovalent $4d$ element Rh$^{4+}$, situated directly above Ir in the periodic table.
The parent Rh compound, Sr$_2$RhO$_4$, is isostructural to Sr$_2$IrO$_4$, but exhibits a smaller
in-plane octahedral rotation angle of $\approx $9.7$^\circ$~\cite{Rh214_1994}. However, Sr$_2$RhO$_4$ is a paramagnetic correlated
metal~\cite{Rh214_1994,Rh_Perry2006, Rh_Moon2006}, characterized by a weaker SOC strength combined with a more effectively screened
Coulomb interaction between O-2$p$ and Rh-4$d$ electrons as compared to Sr$_2$IrO$_4$~\cite{Martins2011}. This is substantiated by a
smaller $U_\text{eff}$ for the Rh $4d$ states, 1.2~eV~\cite{Martins2011}, as compared to $5d$ Ir, 1.6~eV~\cite{Liu2015}; we have used this value for our
DFT+$U$ analysis.
The calculations for Sr$_2$Ir$_{1-x}$Rh$_x$O$_4$ were performed in large supercells of the same size as those used
for the electron doping case, suitable to model the low doping regimes $x$=6.25\% and 12.5\%. The doping induced structural changes are
generally small: we found a slight increase of the tetragonal distortion $c/a$ within the RhO$_6$ octahedron ($c$ and $a$ here refer
to the apical and in-plane Rh-O bond length in the RhO$_6$ sublattice) and a decrease of octahedral rotation angle $\alpha$ compared to
Sr$_2$RhO$_4$ (the structural data are listed in Table.~\ref{tab:Rhstr}).

\begin{table}
\caption{Rh doping ($x=6.25\%$) effect on the structural distortions of
the RhO$_6$ octahedra and its first nearest neighbor (1$nn$) Ir$_{II}$O$_6$ octahedra.
$c$ ($a$) (in $\AA$) represents the apical (in-plane) $M$-O ($M$=Ir or Rh) bond length. $\alpha$ indicates
the in-plane octahedral rotation angle (degree). For comparison, the calculated values for the bulk
Sr$_2$IrO$_4$ and Sr$_2$RhO$_4$ are also given. The Ir$_I$O$_6$ octahedra remain almost unchanged (not listed).
All data were obtained by DFT+$U$+SOC calculations. For Sr$_2$RhO$_4$  we used the experimental lattice parameters
reported in Ref.~\cite{Itoh1995}.
}
\begin{ruledtabular}
\begin{tabular}{lcccc}
         & Sr$_2$IrO$_4$ &  Sr$_2$RhO$_4$   &   \multicolumn{2}{c}{Sr$_2$Ir$_{1-x}$Rh$_x$O$_4$} \\
         \cline{4-5}
         &    IrO$_6$    &    RhO$_6$       &  Ir$_{II}$O$_6$   &    RhO$_6$   \\
\hline
$c$      &     2.071     &     2.067        &     2.062         &    2.104     \\
$a$      &     1.992     &     1.969        &     1.977         &    2.007     \\
$c/a$    &     1.040     &     1.050        &     1.043         &    1.048     \\
$\alpha$ & 13.22$^\circ$ &   11.82$^\circ$   &     12.99$^\circ$ &    12.75$^\circ$     \\
\end{tabular}
\end{ruledtabular}
\label{tab:Rhstr}
\end{table}

\begin{figure}
\begin{center}
\includegraphics[width=0.48\textwidth]{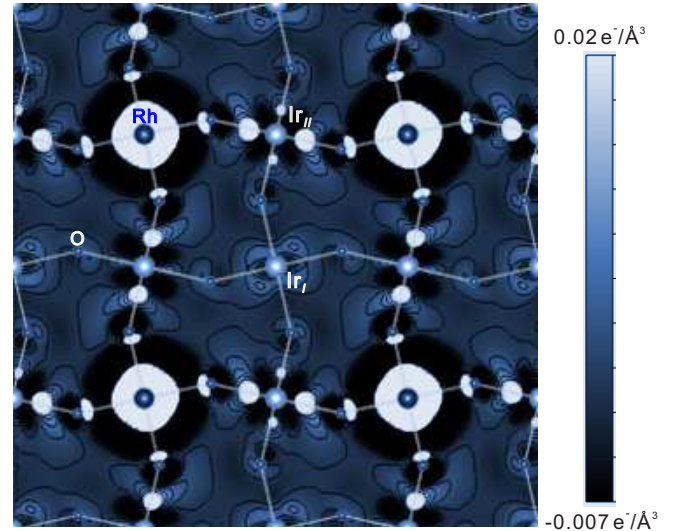}
\end{center}
\caption{(color online)
Charge density difference between the doped $x=6.25\%$ and undoped $x=0$ case within the plane containing the Rh dopant ion.
The gray (blue) intensity scale delineates the charge-transfer process driven by doping which is associated with
an electron transfer from the dark to the white areas. Upon Rh doping, type II iridium ions give away a fraction of
their $t_\text{2g}$ electrons, which are mostly accumulated at the Rh sites, resulting in an effective hole doping.
The Ir-$d$/O-$p$ hybridization along the Rh-O-Ir-O-Rh bond directions is also strongly influenced by doping,
whereas the charge distribution around the Ir$_I$ atoms and around the non-nearest-neighbor O atoms remain almost
unchanged.
}
\label{fig:chg}
\end{figure}

\begin{figure*}
\begin{center}
\includegraphics[width=0.9\textwidth]{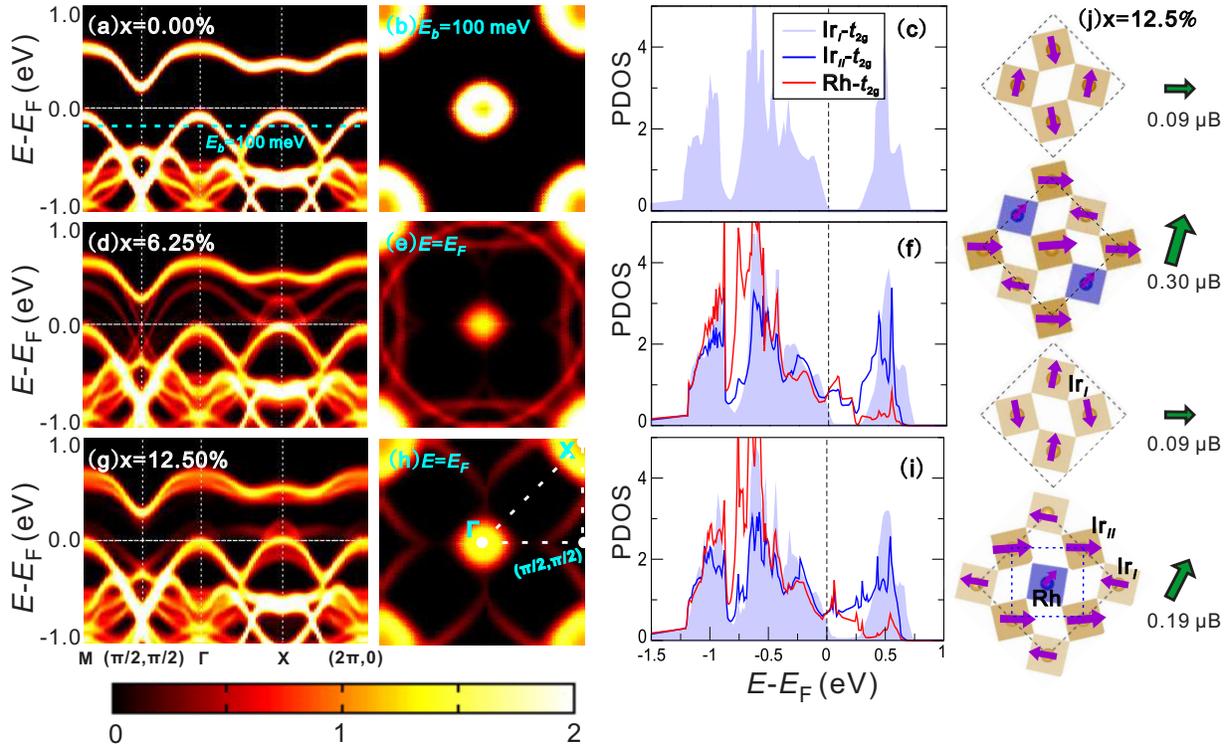}
\end{center}
\caption{(color online) Electronic and magnetic properties of Rh-doped Sr$_2$IrO$_4$.
First column: EBS of (a) undoped and Rh-doped Sr$_2$IrO$_4$ with the nominal hole concentration
(d) $x$=6.25\% and (g) $x$=12.5\%. The color code represents the intensity of the spectral function.
Second column: (b) constant energy contour of undoped  Sr$_2$IrO$_4$ at the binding energy $E_b$=100 meV,
and Fermi surfaces for (e) $x$=6.25\%  and (h) $x$=12.5\%.
Third column:  $t_\text{2g}$ DOS  (states/eV-atom) projected on the Rh, Ir$_I$ and Ir$_{II}$ sites
for (c) $x$ = 0, (f) 6.25\%, and (i) 12.5\%.
Fourth column: schematic plot of the AF-II-like magnetic ordering emerging upon doping
very similar to the ideal AF-II ordering shown in Fig.~\ref{fig:mag}
(given for $x$=12.5\%, a similar ordering is obtained at the lower doping concentration).
The magnitude and direction of the resulting planar magnetic moment obtained by DFT+$U$+SOC is also given.
}
\label{band_Rh}
\end{figure*}

It was proposed that Ir$\rightarrow$Rh chemical substitution leads to hole-transfer from Rh to Ir ideally leading to the formation of
Ir$^{5+}$ and Rh$^{3+}$ ions~\cite{Klein08, Rh_Cao2014, Rh_Clancy2014}. While formally correct, this picture is rather simplified
as it does not consider possible changes in the Ir-$d$/O-$p$ hybridization and does not account for the presence of the two inequivalent
Ir$_I$ and Ir$_{II}$ sites in the compound (see Fig.~\ref{fig:mag}).
To clarify this issue, in Fig. ~\ref{fig:chg} we show our calculated charge density difference between the doped and undoped
sample within the plane containing the substitutional Rh ion. Here the intensity map offers a simple way to visualize the
doping-induced changes in the charge density distribution in terms of charge transfer from the dark to the white regions:
the electron transfer involves an accumulation of electronic
charge around the Rh and (to a lesser extent) O sites and a substantial modification of the Ir-$d$-O-$p$ hybridization, in
particular along the  Rh-O-Ir-O-Rh directions. One can clearly identify the difference between the two kinds of iridium atoms:
Ir$_{II}$ ions, nearest neighbor to Rh, are surrounded by a dark cloud indicating that these atoms donate electron charge to Rh and O,
whereas the charge density around the Ir$_{I}$ ions (all other Ir sites in the supercell) remain essentially unaffected.
This disproportionation between inequivalent Ir sites is also well captured by the DOS shown in Fig.~\ref{band_Rh} (middle panels):
Ir$_{I}$ type atoms preserve a $t_\text{2g}$-projected DOS (shadow areas) almost identical to the Ir atoms in the undoped sample
[Fig.~\ref{band_Rh}(c)], with an essentially insulating character. In contrast, Ir$_{II}$-$t_\text{2g}$ states are pushed above
the Fermi energy  and the peak above the Fermi energy is progressively depleted by increasing doping concentration.

We will  now discuss the changes in the density of states in detail
showing that the transition to the metallic state is caused by the reduced
SOC at the  Rh site as well as a downshift of Rh $4d$ states
compared to the Ir $5d$ states.
Fig. \ref{band_Rh}(f) clearly shows that the shape of the DOS at the Rh atoms closely resembles
that at the Ir atoms, however, the Rh $t_\text{2g}$ states are located at more
negative binding energies than the Ir $t_\text{2g}$ states. This is well understood:
as a result of relativistic effects, the Ir 6$s$ states are closer to the nucleus,
screening it and  pushing the Ir 5$d$ states upwards compared to the Rh $4d$ states;
For isolated atoms the effect is typically 0.5 eV, here, the Rh $4d$ states are shifted
downwards by about 0.2 eV compared to the Ir 5$d$ states. The second
important point is that spin-orbit coupling is much reduced in Rh compared
to Ir. Thus the gap between the lower and upper Hubbard $t_\text{2g}$ band
closes: clearly the $t_\text{2g}$ states that are located above the Fermi-level
for Ir (peak around 0.5 eV) shift into gap shown by the prominent
mid gap Rh $t_\text{2g}$ peak at 0.1 eV. Since Rh has a very small spin orbit
coupling these states even overlap with the lower Hubbard band,
and become partially occupied. Hence the oxidation of the Rh is
Rh$^{4-\delta}$ with $\delta$ being the number of electrons
per site transferred to Rh from the neighboring Ir ions.
An oxidation state of Rh$^{3+}$ is not quite
reached in our calculations, as this would imply a larger
occupation of the Rh upper Hubbard band. However, our DOS data and Bader charge analysis indicate that
$\delta$ is $\approx$ 0.7 $e^-$.
An additional important point is that the neighbouring in-plane Ir$_{II}$ atoms
strongly hybridize with the Rh atom [see Fig.~\ref{band_Rh}(f) and (i)].  As already shown in
Fig. \ref{fig:chg}, each Ir$_{II}$ atom donates the charge $\delta/2$ to the Rh upper Hubbard band.
Hence, Rh creates a fractional hole in the Ir lower Hubbard band,
and the Fermi level is shifted into the lower Hubbard Ir band.
This is in agreement with a recent DFT+$U$ study reporting that Ir$\rightarrow$Rh substitution
is almost isoelectronic and introduces impurity states of predominantly Rh character in the gap region~\cite{Rh_Chikara2015}.
Finally, it is important to note that since Rh-substitution only affects the in-plane Ir
atom, doping results in an essentially two-dimensional (2D) metallicity;
the planes containing only Ir$_I$ atoms retain a predominantly insulating character, very similar to the undoped situation.

The hole doping induced IMT is described in Fig.~\ref{band_Rh} in terms of the EBS, FS, DOS and magnetic ordering.
It can be seen that the insulating state is already perturbed at the lowest doping concentration, due to an upward shift of
the LHB driven by the charge transfer process described above, involving electron transfer from the Ir$_{II}$ sites to the Rh
ions. The LHB crossing at the Fermi energy yields the emergence of hole pockets at $X$ and $\Gamma$ and the formation of
a FS. These results, obtained for fixed (i.e., not doping concentration dependent) values of $U_\text{eff}$ for Rh and Ir,
are generally in good agreement with available ARPES data~\cite{Rh_Cao2014, LaRh2015}; the only exception is
the position of the valence band maxima at $\Gamma$ that is not well described by DFT+$U$ (and $GW$~\cite{liu2016}) as already mentioned
previously. The UHB preserves its fundamental structure upon doping but it looses spectral intensity
and becomes more disperse, suggestive of spectral weight across the gap~\cite{LaRh2015}.
This is also reflected in the weakening of the corresponding DOS with increasing doping [Fig.~\ref{band_Rh}(f) and (i)]
and compatible with the electron transfer process.

We conclude by reporting the impact of Rh doping on the stability of the AF-I magnetic ordering.
Our DFT+$U$+SOC calculations confirm that Rh doping drives a magnetic transition from the AF-I
to AF-II canted ordering in the doping range $0.05 < x < 0.15$~\cite{Rh_Clancy2014, Rh_Ye2015}.
The AF-II state is found to be more stable by 4 meV per formula unit.
 The obtained magnetic state is shown in Fig.~\ref{band_Rh}(j).

\section{Conclusions}\label{sec:sum}

We have investigated the effects of dilute La (electron) and Rh (hole) doping on the electronic, structural and magnetic
properties of Sr$_2$IrO$_4$ by means of magnetically non-collinear DFT+$U$+SOC based supercell calculations combined with
the unfolding band-structure technique. We have shown the effective band structure, Fermi surfaces, density of states, and evolution of the magnetic ordering.

For La doping, our data provide only a partly satisfactory interpretation of the IMT.
Electron doping causes an increase of the chemical potential and drives a moderate renormalization of the electron-electron
interaction quantified by a reduction of the on-site Coulomb interaction $U_\text{eff}$ with doping, from 1.6~eV at $x=0$ to 1.4~eV at $x=12.5\%$.
Upon doping, a metallic state emerges in our simulations, however, the vanishing of the gap between
the lower and upper Hubbard band can not be reproduced satisfactorily. Therefore, some features
are not captured in our simulations, for instance, we fail to observe Dirac like states
at ($\pi/2$,$\pi/2$). Only if $U$ is reduced significantly ($\approx$ 0.5~eV), the Dirac cone emerges as the lower
and upper Hubbard bands approach and the gap closes, in accordance with
previous model studies done at $U=0$~\cite{La_Torre2015}. From a first principles perspective this
is certainly not satisfactory, implying that the one-electron methods used here
are not sufficient to reproduce the experiments satisfactorily.
As for the magnetic properties, although the magnitude of the local moment is only marginally affected by doping,
the characteristic canted AF-I state is locally perturbed in the vicinity of the La site and for doping
concentrations $x>12.5\%$ a regular long-range canted AF-I pattern is lost.

For the Rh doping case our data demonstrate that Rh doping is responsible for a charge redistribution involving
predominantly  a fractional electron transfer to the Rh sites from the nearest neighbor Ir sites (Ir$_{II}$).
This in turn leads to the formation of two dimensional metallic Rh-Ir$_{II}$ planes intercalated by unperturbed insulating Ir$_I$ planes.
The emergence of this metallic state is assisted by the smaller SOC strength on the Rh site: the DOS indicates that
the  Rh $t_\text{2g}$ states are located at more negative binding energies than the corresponding Ir states mostly
due to smaller relativistic effects.
As a consequence, since the upper Hubbard Rh-Ir-$t_{2g}$ states are  located below the Fermi-level,
the lower Hubbard band looses electrons and becomes partially occupied.
Upon Rh doping the canted AF-I state undergoes a substantial modification, manifested by a
the flipping of the net in-plane FM moment in the Rh-Ir$_{II}$ planes and
lifts the magnetic ordering from AF-I to AF-II type.
All our results are in generally good agreement with available ARPES and neutron measurements.

\begin{acknowledgments}
This work was supported by China Scholarship Council (CSC)-Austrian Science Fund (FWF) Scholarship Program,
by the joint FWF and Indian Department of Science and Technology (DST) project INDOX (I1490-N19), and
by the FWF-SFB ViCoM (Grant No. F41). Computing time at the Vienna Scientific Cluster is greatly acknowledged.
\end{acknowledgments}

\bibliographystyle{apsrev}
\bibliography{reference}

\begin{thebibliography}{43}
\expandafter\ifx\csname natexlab\endcsname\relax\def\natexlab#1{#1}\fi
\expandafter\ifx\csname bibnamefont\endcsname\relax
  \def\bibnamefont#1{#1}\fi
\expandafter\ifx\csname bibfnamefont\endcsname\relax
  \def\bibfnamefont#1{#1}\fi
\expandafter\ifx\csname citenamefont\endcsname\relax
  \def\citenamefont#1{#1}\fi
\expandafter\ifx\csname url\endcsname\relax
  \def\url#1{\texttt{#1}}\fi
\expandafter\ifx\csname urlprefix\endcsname\relax\def\urlprefix{URL }\fi
\providecommand{\bibinfo}[2]{#2}
\providecommand{\eprint}[2][]{\url{#2}}

\bibitem[{\citenamefont{Kim et~al.}(2008)\citenamefont{Kim, Jin, Moon, Kim,
  Park, Leem, Yu, Noh, Kim, Oh et~al.}}]{Kim2008}
\bibinfo{author}{\bibfnamefont{B.~J.} \bibnamefont{Kim}},
  \bibinfo{author}{\bibfnamefont{H.}~\bibnamefont{Jin}},
  \bibinfo{author}{\bibfnamefont{S.}~\bibnamefont{Moon}},
  \bibinfo{author}{\bibfnamefont{J.~Y.} \bibnamefont{Kim}},
  \bibinfo{author}{\bibfnamefont{B.~G.} \bibnamefont{Park}},
  \bibinfo{author}{\bibfnamefont{C.}~\bibnamefont{Leem}},
  \bibinfo{author}{\bibfnamefont{J.}~\bibnamefont{Yu}},
  \bibinfo{author}{\bibfnamefont{T.}~\bibnamefont{Noh}},
  \bibinfo{author}{\bibfnamefont{C.}~\bibnamefont{Kim}},
  \bibinfo{author}{\bibfnamefont{S.~J.} \bibnamefont{Oh}},
  \bibnamefont{et~al.}, \bibinfo{journal}{Phys. Rev. Lett.}
  \textbf{\bibinfo{volume}{101}}, \bibinfo{pages}{076402}
  (\bibinfo{year}{2008}).

\bibitem[{\citenamefont{Kim et~al.}(2009)\citenamefont{Kim, Ohsumi, Komesu,
  Sakai, Morita, Takagi, and Arima}}]{Kim2009}
\bibinfo{author}{\bibfnamefont{B.~J.} \bibnamefont{Kim}},
  \bibinfo{author}{\bibfnamefont{H.}~\bibnamefont{Ohsumi}},
  \bibinfo{author}{\bibfnamefont{T.}~\bibnamefont{Komesu}},
  \bibinfo{author}{\bibfnamefont{S.}~\bibnamefont{Sakai}},
  \bibinfo{author}{\bibfnamefont{T.}~\bibnamefont{Morita}},
  \bibinfo{author}{\bibfnamefont{H.}~\bibnamefont{Takagi}}, \bibnamefont{and}
  \bibinfo{author}{\bibfnamefont{T.}~\bibnamefont{Arima}},
  \bibinfo{journal}{Science} \textbf{\bibinfo{volume}{323}},
  \bibinfo{pages}{1329} (\bibinfo{year}{2009}).

\bibitem[{\citenamefont{Jackeli and Khaliullin}(2009)}]{Jackeli2009}
\bibinfo{author}{\bibfnamefont{G.}~\bibnamefont{Jackeli}} \bibnamefont{and}
  \bibinfo{author}{\bibfnamefont{G.}~\bibnamefont{Khaliullin}},
  \bibinfo{journal}{Phys. Rev. Lett.} \textbf{\bibinfo{volume}{102}},
  \bibinfo{pages}{017205} (\bibinfo{year}{2009}).

\bibitem[{\citenamefont{Kim et~al.}(2012{\natexlab{a}})\citenamefont{Kim, Choi,
  Kim, Mitchell, G.~Jackeli, van~den Brink, Khaliullin, and
  Kim}}]{Khaliullin2012}
\bibinfo{author}{\bibfnamefont{J.~W.} \bibnamefont{Kim}},
  \bibinfo{author}{\bibfnamefont{Y.}~\bibnamefont{Choi}},
  \bibinfo{author}{\bibfnamefont{J.}~\bibnamefont{Kim}},
  \bibinfo{author}{\bibfnamefont{J.~F.} \bibnamefont{Mitchell}},
  \bibinfo{author}{\bibfnamefont{M.~D.} \bibnamefont{G.~Jackeli}},
  \bibinfo{author}{\bibfnamefont{J.}~\bibnamefont{van~den Brink}},
  \bibinfo{author}{\bibfnamefont{G.}~\bibnamefont{Khaliullin}},
  \bibnamefont{and} \bibinfo{author}{\bibfnamefont{B.~J.} \bibnamefont{Kim}},
  \bibinfo{journal}{Phys. Rev. Lett.} \textbf{\bibinfo{volume}{109}},
  \bibinfo{pages}{037204} (\bibinfo{year}{2012}{\natexlab{a}}).

\bibitem[{\citenamefont{Kim et~al.}(2014{\natexlab{a}})\citenamefont{Kim,
  Daghofer, Said, Gog, van~den Brink, Khaliullin, and Kim}}]{Kim2014exciton}
\bibinfo{author}{\bibfnamefont{J.}~\bibnamefont{Kim}},
  \bibinfo{author}{\bibfnamefont{M.}~\bibnamefont{Daghofer}},
  \bibinfo{author}{\bibfnamefont{A.~H.} \bibnamefont{Said}},
  \bibinfo{author}{\bibfnamefont{T.}~\bibnamefont{Gog}},
  \bibinfo{author}{\bibfnamefont{J.}~\bibnamefont{van~den Brink}},
  \bibinfo{author}{\bibfnamefont{G.}~\bibnamefont{Khaliullin}},
  \bibnamefont{and} \bibinfo{author}{\bibfnamefont{B.~J.} \bibnamefont{Kim}},
  \bibinfo{journal}{Nature communications} \textbf{\bibinfo{volume}{5}},
  \bibinfo{pages}{4453} (\bibinfo{year}{2014}{\natexlab{a}}).

\bibitem[{\citenamefont{Liu et~al.}(2015)\citenamefont{Liu, Khmelevskyi, Kim,
  Marsman, Li, Chen, Sarma, Kresse, and Franchini}}]{Liu2015}
\bibinfo{author}{\bibfnamefont{P.}~\bibnamefont{Liu}},
  \bibinfo{author}{\bibfnamefont{S.}~\bibnamefont{Khmelevskyi}},
  \bibinfo{author}{\bibfnamefont{B.}~\bibnamefont{Kim}},
  \bibinfo{author}{\bibfnamefont{M.}~\bibnamefont{Marsman}},
  \bibinfo{author}{\bibfnamefont{D.}~\bibnamefont{Li}},
  \bibinfo{author}{\bibfnamefont{X.-Q.} \bibnamefont{Chen}},
  \bibinfo{author}{\bibfnamefont{D.~D.} \bibnamefont{Sarma}},
  \bibinfo{author}{\bibfnamefont{G.}~\bibnamefont{Kresse}}, \bibnamefont{and}
  \bibinfo{author}{\bibfnamefont{C.}~\bibnamefont{Franchini}},
  \bibinfo{journal}{Phys. Rev. B} \textbf{\bibinfo{volume}{92}},
  \bibinfo{pages}{054428} (\bibinfo{year}{2015}).

\bibitem[{\citenamefont{Hou et~al.}(2016)\citenamefont{Hou, Xiang, and
  Gong}}]{Hou2016}
\bibinfo{author}{\bibfnamefont{Y.}~\bibnamefont{Hou}},
  \bibinfo{author}{\bibfnamefont{H.}~\bibnamefont{Xiang}}, \bibnamefont{and}
  \bibinfo{author}{\bibfnamefont{X.}~\bibnamefont{Gong}}, \bibinfo{journal}{New
  J. Phys.} \textbf{\bibinfo{volume}{18}}, \bibinfo{pages}{043007}
  (\bibinfo{year}{2016}).

\bibitem[{\citenamefont{Kim et~al.}(2012{\natexlab{b}})\citenamefont{Kim, Casa,
  Upton, Gog, Kim, Mitchell, van Veenendaal, Daghofer, van~den Brink,
  Khaliullin et~al.}}]{cuprate-1}
\bibinfo{author}{\bibfnamefont{J.}~\bibnamefont{Kim}},
  \bibinfo{author}{\bibfnamefont{D.}~\bibnamefont{Casa}},
  \bibinfo{author}{\bibfnamefont{M.~H.} \bibnamefont{Upton}},
  \bibinfo{author}{\bibfnamefont{T.}~\bibnamefont{Gog}},
  \bibinfo{author}{\bibfnamefont{Y.-J.} \bibnamefont{Kim}},
  \bibinfo{author}{\bibfnamefont{J.~F.} \bibnamefont{Mitchell}},
  \bibinfo{author}{\bibfnamefont{M.}~\bibnamefont{van Veenendaal}},
  \bibinfo{author}{\bibfnamefont{M.}~\bibnamefont{Daghofer}},
  \bibinfo{author}{\bibfnamefont{J.}~\bibnamefont{van~den Brink}},
  \bibinfo{author}{\bibfnamefont{G.}~\bibnamefont{Khaliullin}},
  \bibnamefont{et~al.}, \bibinfo{journal}{Phys. Rev. Lett.}
  \textbf{\bibinfo{volume}{108}}, \bibinfo{pages}{177003}
  (\bibinfo{year}{2012}{\natexlab{b}}).

\bibitem[{\citenamefont{Fujiyama et~al.}(2012)\citenamefont{Fujiyama, Ohsumi,
  Komesu, Matsuno, Kim, Takata, Arima, and Takagi}}]{FujiyamaPRL2012}
\bibinfo{author}{\bibfnamefont{S.}~\bibnamefont{Fujiyama}},
  \bibinfo{author}{\bibfnamefont{H.}~\bibnamefont{Ohsumi}},
  \bibinfo{author}{\bibfnamefont{T.}~\bibnamefont{Komesu}},
  \bibinfo{author}{\bibfnamefont{J.}~\bibnamefont{Matsuno}},
  \bibinfo{author}{\bibfnamefont{B.~J.} \bibnamefont{Kim}},
  \bibinfo{author}{\bibfnamefont{M.}~\bibnamefont{Takata}},
  \bibinfo{author}{\bibfnamefont{T.}~\bibnamefont{Arima}}, \bibnamefont{and}
  \bibinfo{author}{\bibfnamefont{H.}~\bibnamefont{Takagi}},
  \bibinfo{journal}{Phys. Rev. Lett.} \textbf{\bibinfo{volume}{108}},
  \bibinfo{pages}{247212} (\bibinfo{year}{2012}).

\bibitem[{\citenamefont{Wang and Senthil}(2011)}]{Wang2011}
\bibinfo{author}{\bibfnamefont{F.}~\bibnamefont{Wang}} \bibnamefont{and}
  \bibinfo{author}{\bibfnamefont{T.}~\bibnamefont{Senthil}},
  \bibinfo{journal}{Phys. Rev. Lett.} \textbf{\bibinfo{volume}{106}},
  \bibinfo{pages}{136402} (\bibinfo{year}{2011}).

\bibitem[{\citenamefont{Watanabe et~al.}(2013)\citenamefont{Watanabe,
  Shirakawa, and Yunoki}}]{Watanabe2013}
\bibinfo{author}{\bibfnamefont{H.}~\bibnamefont{Watanabe}},
  \bibinfo{author}{\bibfnamefont{T.}~\bibnamefont{Shirakawa}},
  \bibnamefont{and} \bibinfo{author}{\bibfnamefont{S.}~\bibnamefont{Yunoki}},
  \bibinfo{journal}{Phys. Rev. Lett.} \textbf{\bibinfo{volume}{110}},
  \bibinfo{pages}{027002} (\bibinfo{year}{2013}).

\bibitem[{\citenamefont{Kim et~al.}(2014{\natexlab{b}})\citenamefont{Kim,
  Krupin, Denlinger, Bostwick, Rotenberg, Zhao, Mitchell, Allen, and
  Kim}}]{KimScience2014}
\bibinfo{author}{\bibfnamefont{Y.~K.} \bibnamefont{Kim}},
  \bibinfo{author}{\bibfnamefont{O.}~\bibnamefont{Krupin}},
  \bibinfo{author}{\bibfnamefont{J.~D.} \bibnamefont{Denlinger}},
  \bibinfo{author}{\bibfnamefont{A.}~\bibnamefont{Bostwick}},
  \bibinfo{author}{\bibfnamefont{E.}~\bibnamefont{Rotenberg}},
  \bibinfo{author}{\bibfnamefont{Q.}~\bibnamefont{Zhao}},
  \bibinfo{author}{\bibfnamefont{J.~F.} \bibnamefont{Mitchell}},
  \bibinfo{author}{\bibfnamefont{J.~W.} \bibnamefont{Allen}}, \bibnamefont{and}
  \bibinfo{author}{\bibfnamefont{B.~J.} \bibnamefont{Kim}},
  \bibinfo{journal}{Science} \textbf{\bibinfo{volume}{345}},
  \bibinfo{pages}{187} (\bibinfo{year}{2014}{\natexlab{b}}).

\bibitem[{\citenamefont{Yan et~al.}(2015)\citenamefont{Yan, Ren, Xu, Xie, Tao,
  Choi, Lee, Choi, Zhang, and Feng}}]{Yan2015}
\bibinfo{author}{\bibfnamefont{Y.~J.} \bibnamefont{Yan}},
  \bibinfo{author}{\bibfnamefont{M.~Q.} \bibnamefont{Ren}},
  \bibinfo{author}{\bibfnamefont{H.~C.} \bibnamefont{Xu}},
  \bibinfo{author}{\bibfnamefont{B.~P.} \bibnamefont{Xie}},
  \bibinfo{author}{\bibfnamefont{R.}~\bibnamefont{Tao}},
  \bibinfo{author}{\bibfnamefont{H.~Y.} \bibnamefont{Choi}},
  \bibinfo{author}{\bibfnamefont{N.}~\bibnamefont{Lee}},
  \bibinfo{author}{\bibfnamefont{Y.~J.} \bibnamefont{Choi}},
  \bibinfo{author}{\bibfnamefont{T.}~\bibnamefont{Zhang}}, \bibnamefont{and}
  \bibinfo{author}{\bibfnamefont{D.~L.} \bibnamefont{Feng}},
  \bibinfo{journal}{Phys. Rev. X} \textbf{\bibinfo{volume}{5}},
  \bibinfo{pages}{041018} (\bibinfo{year}{2015}).

\bibitem[{\citenamefont{Kim et~al.}(2015)\citenamefont{Kim, Sung, Denlinger,
  and Kim}}]{Kim_n2015}
\bibinfo{author}{\bibfnamefont{Y.~K.} \bibnamefont{Kim}},
  \bibinfo{author}{\bibfnamefont{N.~H.} \bibnamefont{Sung}},
  \bibinfo{author}{\bibfnamefont{J.~D.} \bibnamefont{Denlinger}},
  \bibnamefont{and} \bibinfo{author}{\bibfnamefont{B.~J.} \bibnamefont{Kim}},
  \bibinfo{journal}{Nature Physics} \textbf{\bibinfo{volume}{12}},
  \bibinfo{pages}{37} (\bibinfo{year}{2015}).

\bibitem[{\citenamefont{de~la Torre et~al.}(2015)\citenamefont{de~la Torre,
  Walker, Bruno, Ricc\'o, Wang, Lezama, Scheerer, Giriat, Jaccard, Berthod
  et~al.}}]{La_Torre2015}
\bibinfo{author}{\bibfnamefont{A.}~\bibnamefont{de~la Torre}},
  \bibinfo{author}{\bibfnamefont{S.~M.} \bibnamefont{Walker}},
  \bibinfo{author}{\bibfnamefont{F.~Y.} \bibnamefont{Bruno}},
  \bibinfo{author}{\bibfnamefont{S.}~\bibnamefont{Ricc\'o}},
  \bibinfo{author}{\bibfnamefont{Z.}~\bibnamefont{Wang}},
  \bibinfo{author}{\bibfnamefont{I.~G.} \bibnamefont{Lezama}},
  \bibinfo{author}{\bibfnamefont{G.}~\bibnamefont{Scheerer}},
  \bibinfo{author}{\bibfnamefont{G.}~\bibnamefont{Giriat}},
  \bibinfo{author}{\bibfnamefont{D.}~\bibnamefont{Jaccard}},
  \bibinfo{author}{\bibfnamefont{C.}~\bibnamefont{Berthod}},
  \bibnamefont{et~al.}, \bibinfo{journal}{Phys. Rev. Lett.}
  \textbf{\bibinfo{volume}{115}}, \bibinfo{pages}{176402}
  (\bibinfo{year}{2015}).

\bibitem[{\citenamefont{Lee et~al.}(2006)\citenamefont{Lee, Nagaosa, and
  Wen}}]{Lee2006}
\bibinfo{author}{\bibfnamefont{P.~A.} \bibnamefont{Lee}},
  \bibinfo{author}{\bibfnamefont{N.}~\bibnamefont{Nagaosa}}, \bibnamefont{and}
  \bibinfo{author}{\bibfnamefont{X.-G.} \bibnamefont{Wen}},
  \bibinfo{journal}{Rev. Mod. Phys.} \textbf{\bibinfo{volume}{78}},
  \bibinfo{pages}{17} (\bibinfo{year}{2006}).

\bibitem[{\citenamefont{Klein and Terasaki}(2008)}]{Klein08}
\bibinfo{author}{\bibfnamefont{Y.}~\bibnamefont{Klein}} \bibnamefont{and}
  \bibinfo{author}{\bibfnamefont{I.}~\bibnamefont{Terasaki}},
  \bibinfo{journal}{J. Phys.: Condens. Matter} \textbf{\bibinfo{volume}{20}},
  \bibinfo{pages}{295201} (\bibinfo{year}{2008}).

\bibitem[{\citenamefont{Ge et~al.}(2011)\citenamefont{Ge, Qi, Korneta, Long,
  Schlottmann, Crummett, and Cao}}]{La_Ge2011}
\bibinfo{author}{\bibfnamefont{M.}~\bibnamefont{Ge}},
  \bibinfo{author}{\bibfnamefont{T.~F.} \bibnamefont{Qi}},
  \bibinfo{author}{\bibfnamefont{O.~B.} \bibnamefont{Korneta}},
  \bibinfo{author}{\bibfnamefont{D.~E.~D.} \bibnamefont{Long}},
  \bibinfo{author}{\bibfnamefont{P.}~\bibnamefont{Schlottmann}},
  \bibinfo{author}{\bibfnamefont{W.~P.} \bibnamefont{Crummett}},
  \bibnamefont{and} \bibinfo{author}{\bibfnamefont{G.}~\bibnamefont{Cao}},
  \bibinfo{journal}{Phys. Rev. B} \textbf{\bibinfo{volume}{84}},
  \bibinfo{pages}{100402(R)} (\bibinfo{year}{2011}).

\bibitem[{\citenamefont{Lee et~al.}(2012)\citenamefont{Lee, Krockenberger,
  Takahashi, Kawasaki, and Tokura}}]{All_Lee2012}
\bibinfo{author}{\bibfnamefont{J.~S.} \bibnamefont{Lee}},
  \bibinfo{author}{\bibfnamefont{Y.}~\bibnamefont{Krockenberger}},
  \bibinfo{author}{\bibfnamefont{K.~S.} \bibnamefont{Takahashi}},
  \bibinfo{author}{\bibfnamefont{M.}~\bibnamefont{Kawasaki}}, \bibnamefont{and}
  \bibinfo{author}{\bibfnamefont{Y.}~\bibnamefont{Tokura}},
  \bibinfo{journal}{Phys. Rev. B} \textbf{\bibinfo{volume}{85}},
  \bibinfo{pages}{035101} (\bibinfo{year}{2012}).

\bibitem[{\citenamefont{Qi et~al.}(2012)\citenamefont{Qi, Korneta, Li,
  Butrouna, Cao, Wan, Schlottmann, Kaul, and Cao}}]{Rh_Qi2012}
\bibinfo{author}{\bibfnamefont{T.~F.} \bibnamefont{Qi}},
  \bibinfo{author}{\bibfnamefont{O.~B.} \bibnamefont{Korneta}},
  \bibinfo{author}{\bibfnamefont{L.}~\bibnamefont{Li}},
  \bibinfo{author}{\bibfnamefont{K.}~\bibnamefont{Butrouna}},
  \bibinfo{author}{\bibfnamefont{V.~S.} \bibnamefont{Cao}},
  \bibinfo{author}{\bibfnamefont{X.}~\bibnamefont{Wan}},
  \bibinfo{author}{\bibfnamefont{P.}~\bibnamefont{Schlottmann}},
  \bibinfo{author}{\bibfnamefont{R.~K.} \bibnamefont{Kaul}}, \bibnamefont{and}
  \bibinfo{author}{\bibfnamefont{G.}~\bibnamefont{Cao}},
  \bibinfo{journal}{Phys. Rev. B} \textbf{\bibinfo{volume}{86}},
  \bibinfo{pages}{125105} (\bibinfo{year}{2012}).

\bibitem[{\citenamefont{Cao et~al.}(2016)\citenamefont{Cao, Wang, Waugh, Reber,
  Li, Zhou, Parham, Plumb, Rotenberg, Bostwick et~al.}}]{Rh_Cao2014}
\bibinfo{author}{\bibfnamefont{Y.}~\bibnamefont{Cao}},
  \bibinfo{author}{\bibfnamefont{Q.}~\bibnamefont{Wang}},
  \bibinfo{author}{\bibfnamefont{J.~A.} \bibnamefont{Waugh}},
  \bibinfo{author}{\bibfnamefont{T.~J.} \bibnamefont{Reber}},
  \bibinfo{author}{\bibfnamefont{H.}~\bibnamefont{Li}},
  \bibinfo{author}{\bibfnamefont{X.}~\bibnamefont{Zhou}},
  \bibinfo{author}{\bibfnamefont{S.}~\bibnamefont{Parham}},
  \bibinfo{author}{\bibfnamefont{N.~C.} \bibnamefont{Plumb}},
  \bibinfo{author}{\bibfnamefont{E.}~\bibnamefont{Rotenberg}},
  \bibinfo{author}{\bibfnamefont{A.}~\bibnamefont{Bostwick}},
  \bibnamefont{et~al.}, \bibinfo{journal}{Nat. Comm.}
  \textbf{\bibinfo{volume}{7}}, \bibinfo{pages}{11367} (\bibinfo{year}{2016}).

\bibitem[{\citenamefont{Clancy et~al.}(2014)\citenamefont{Clancy, Lupascu,
  Gretarsson, Islam, Hu, Casa, Nelson, LaMarra, Cao, and Kim}}]{Rh_Clancy2014}
\bibinfo{author}{\bibfnamefont{J.~P.} \bibnamefont{Clancy}},
  \bibinfo{author}{\bibfnamefont{A.}~\bibnamefont{Lupascu}},
  \bibinfo{author}{\bibfnamefont{H.}~\bibnamefont{Gretarsson}},
  \bibinfo{author}{\bibfnamefont{Z.}~\bibnamefont{Islam}},
  \bibinfo{author}{\bibfnamefont{Y.~F.} \bibnamefont{Hu}},
  \bibinfo{author}{\bibfnamefont{D.}~\bibnamefont{Casa}},
  \bibinfo{author}{\bibfnamefont{C.~S.} \bibnamefont{Nelson}},
  \bibinfo{author}{\bibfnamefont{S.~C.} \bibnamefont{LaMarra}},
  \bibinfo{author}{\bibfnamefont{G.}~\bibnamefont{Cao}}, \bibnamefont{and}
  \bibinfo{author}{\bibfnamefont{Y.-J.} \bibnamefont{Kim}},
  \bibinfo{journal}{Phys. Rev. B} \textbf{\bibinfo{volume}{89}},
  \bibinfo{pages}{054409} (\bibinfo{year}{2014}).

\bibitem[{\citenamefont{Ye et~al.}(2015)\citenamefont{Ye, Wang, Hoffmann, Wang,
  Chi, Matsuda, Chakoumakos, Fernandez-Baca, and Cao}}]{Rh_Ye2015}
\bibinfo{author}{\bibfnamefont{F.}~\bibnamefont{Ye}},
  \bibinfo{author}{\bibfnamefont{X.}~\bibnamefont{Wang}},
  \bibinfo{author}{\bibfnamefont{C.}~\bibnamefont{Hoffmann}},
  \bibinfo{author}{\bibfnamefont{J.}~\bibnamefont{Wang}},
  \bibinfo{author}{\bibfnamefont{S.}~\bibnamefont{Chi}},
  \bibinfo{author}{\bibfnamefont{M.}~\bibnamefont{Matsuda}},
  \bibinfo{author}{\bibfnamefont{B.~C.} \bibnamefont{Chakoumakos}},
  \bibinfo{author}{\bibfnamefont{J.~A.} \bibnamefont{Fernandez-Baca}},
  \bibnamefont{and} \bibinfo{author}{\bibfnamefont{G.}~\bibnamefont{Cao}},
  \bibinfo{journal}{Phys. Rev. B} \textbf{\bibinfo{volume}{92}},
  \bibinfo{pages}{201112(R)} (\bibinfo{year}{2015}).

\bibitem[{\citenamefont{Chen et~al.}(2015)\citenamefont{Chen, Hogan, Walkup,
  Zhou, Pokharel, Yao, Tian, Ward, Zhao, Parshall et~al.}}]{La_Chen2015}
\bibinfo{author}{\bibfnamefont{X.}~\bibnamefont{Chen}},
  \bibinfo{author}{\bibfnamefont{T.}~\bibnamefont{Hogan}},
  \bibinfo{author}{\bibfnamefont{D.}~\bibnamefont{Walkup}},
  \bibinfo{author}{\bibfnamefont{W.}~\bibnamefont{Zhou}},
  \bibinfo{author}{\bibfnamefont{M.}~\bibnamefont{Pokharel}},
  \bibinfo{author}{\bibfnamefont{M.}~\bibnamefont{Yao}},
  \bibinfo{author}{\bibfnamefont{W.}~\bibnamefont{Tian}},
  \bibinfo{author}{\bibfnamefont{T.~Z.} \bibnamefont{Ward}},
  \bibinfo{author}{\bibfnamefont{Y.}~\bibnamefont{Zhao}},
  \bibinfo{author}{\bibfnamefont{D.}~\bibnamefont{Parshall}},
  \bibnamefont{et~al.}, \bibinfo{journal}{Phys. Rev. B}
  \textbf{\bibinfo{volume}{92}}, \bibinfo{pages}{075125}
  (\bibinfo{year}{2015}).

\bibitem[{\citenamefont{Brouet et~al.}(2015)\citenamefont{Brouet, Mansart,
  Perfetti, Piovera, Vobornik, F\'evre, Bertran, Riggs, Shapiro, Giraldo-Gallo
  et~al.}}]{LaRh2015}
\bibinfo{author}{\bibfnamefont{V.}~\bibnamefont{Brouet}},
  \bibinfo{author}{\bibfnamefont{J.}~\bibnamefont{Mansart}},
  \bibinfo{author}{\bibfnamefont{L.}~\bibnamefont{Perfetti}},
  \bibinfo{author}{\bibfnamefont{C.}~\bibnamefont{Piovera}},
  \bibinfo{author}{\bibfnamefont{I.}~\bibnamefont{Vobornik}},
  \bibinfo{author}{\bibfnamefont{P.~L.} \bibnamefont{F\'evre}},
  \bibinfo{author}{\bibfnamefont{F.}~\bibnamefont{Bertran}},
  \bibinfo{author}{\bibfnamefont{S.~C.} \bibnamefont{Riggs}},
  \bibinfo{author}{\bibfnamefont{M.~C.} \bibnamefont{Shapiro}},
  \bibinfo{author}{\bibfnamefont{P.}~\bibnamefont{Giraldo-Gallo}},
  \bibnamefont{et~al.}, \bibinfo{journal}{Phys. Rev. B}
  \textbf{\bibinfo{volume}{92}}, \bibinfo{pages}{081117}
  (\bibinfo{year}{2015}).

\bibitem[{\citenamefont{Calder et~al.}(2015)\citenamefont{Calder, Kim, Cao,
  Cantoni, May, Cao, Aczel, Matsuda, Choi, Haskel et~al.}}]{Ru_Calder2015}
\bibinfo{author}{\bibfnamefont{S.}~\bibnamefont{Calder}},
  \bibinfo{author}{\bibfnamefont{J.~W.} \bibnamefont{Kim}},
  \bibinfo{author}{\bibfnamefont{G.~X.} \bibnamefont{Cao}},
  \bibinfo{author}{\bibfnamefont{C.}~\bibnamefont{Cantoni}},
  \bibinfo{author}{\bibfnamefont{A.~F.} \bibnamefont{May}},
  \bibinfo{author}{\bibfnamefont{H.~B.} \bibnamefont{Cao}},
  \bibinfo{author}{\bibfnamefont{A.~A.} \bibnamefont{Aczel}},
  \bibinfo{author}{\bibfnamefont{M.}~\bibnamefont{Matsuda}},
  \bibinfo{author}{\bibfnamefont{Y.}~\bibnamefont{Choi}},
  \bibinfo{author}{\bibfnamefont{D.}~\bibnamefont{Haskel}},
  \bibnamefont{et~al.}, \bibinfo{journal}{Phys. Rev. B}
  \textbf{\bibinfo{volume}{92}}, \bibinfo{pages}{165128}
  (\bibinfo{year}{2015}).

\bibitem[{\citenamefont{Popescu and Zunger}(2010)}]{Popescu2010}
\bibinfo{author}{\bibfnamefont{V.}~\bibnamefont{Popescu}} \bibnamefont{and}
  \bibinfo{author}{\bibfnamefont{A.}~\bibnamefont{Zunger}},
  \bibinfo{journal}{Phys. Rev. Lett.} \textbf{\bibinfo{volume}{104}},
  \bibinfo{pages}{236403} (\bibinfo{year}{2010}).

\bibitem[{\citenamefont{Popescu and Zunger}(2012)}]{Popescu2012}
\bibinfo{author}{\bibfnamefont{V.}~\bibnamefont{Popescu}} \bibnamefont{and}
  \bibinfo{author}{\bibfnamefont{A.}~\bibnamefont{Zunger}},
  \bibinfo{journal}{Phys. Rev. B} \textbf{\bibinfo{volume}{85}},
  \bibinfo{pages}{085201} (\bibinfo{year}{2012}).

\bibitem[{\citenamefont{Eckhardt et~al.}(2014)\citenamefont{Eckhardt, Hummer,
  and Kresse}}]{Eckhardt2014}
\bibinfo{author}{\bibfnamefont{C.}~\bibnamefont{Eckhardt}},
  \bibinfo{author}{\bibfnamefont{K.}~\bibnamefont{Hummer}}, \bibnamefont{and}
  \bibinfo{author}{\bibfnamefont{G.}~\bibnamefont{Kresse}},
  \bibinfo{journal}{Phys. Rev. B} \textbf{\bibinfo{volume}{89}},
  \bibinfo{pages}{165201} (\bibinfo{year}{2014}).

\bibitem[{\citenamefont{Reticcioli et~al.}(2016)\citenamefont{Reticcioli,
  Profeta, Franchini, and Continenza}}]{Michele2015}
\bibinfo{author}{\bibfnamefont{M.}~\bibnamefont{Reticcioli}},
  \bibinfo{author}{\bibfnamefont{G.}~\bibnamefont{Profeta}},
  \bibinfo{author}{\bibfnamefont{C.}~\bibnamefont{Franchini}},
  \bibnamefont{and}
  \bibinfo{author}{\bibfnamefont{A.}~\bibnamefont{Continenza}},
  \bibinfo{journal}{Journal of Physics: Conference Series}
  \textbf{\bibinfo{volume}{689}}, \bibinfo{pages}{012027}
  (\bibinfo{year}{2016}).

\bibitem[{\citenamefont{Bl\"{o}chl}(1994)}]{PAW}
\bibinfo{author}{\bibfnamefont{P.~E.} \bibnamefont{Bl\"{o}chl}},
  \bibinfo{journal}{Phys. Rev. B} \textbf{\bibinfo{volume}{50}},
  \bibinfo{pages}{17953} (\bibinfo{year}{1994}).

\bibitem[{\citenamefont{Kresse and Hafner}(1993)}]{Kresse-1}
\bibinfo{author}{\bibfnamefont{G.}~\bibnamefont{Kresse}} \bibnamefont{and}
  \bibinfo{author}{\bibfnamefont{J.}~\bibnamefont{Hafner}},
  \bibinfo{journal}{Phys. Rev. B} \textbf{\bibinfo{volume}{47}},
  \bibinfo{pages}{558} (\bibinfo{year}{1993}).

\bibitem[{\citenamefont{Kresse and Furthm\"{u}ller}(1996)}]{Kresse-2}
\bibinfo{author}{\bibfnamefont{G.}~\bibnamefont{Kresse}} \bibnamefont{and}
  \bibinfo{author}{\bibfnamefont{J.}~\bibnamefont{Furthm\"{u}ller}},
  \bibinfo{journal}{Phys. Rev. B} \textbf{\bibinfo{volume}{54}},
  \bibinfo{pages}{11169} (\bibinfo{year}{1996}).

\bibitem[{\citenamefont{Perdew et~al.}(1996)\citenamefont{Perdew, Burke, and
  Ernzerhof}}]{PBE}
\bibinfo{author}{\bibfnamefont{J.~P.} \bibnamefont{Perdew}},
  \bibinfo{author}{\bibfnamefont{K.}~\bibnamefont{Burke}}, \bibnamefont{and}
  \bibinfo{author}{\bibfnamefont{M.}~\bibnamefont{Ernzerhof}},
  \bibinfo{journal}{Phys. Rev. Lett.} \textbf{\bibinfo{volume}{77}},
  \bibinfo{pages}{3865} (\bibinfo{year}{1996}).

\bibitem[{\citenamefont{Martins et~al.}(2011)\citenamefont{Martins, Aichhorn,
  Vaugier, and Biermann}}]{Martins2011}
\bibinfo{author}{\bibfnamefont{C.}~\bibnamefont{Martins}},
  \bibinfo{author}{\bibfnamefont{M.}~\bibnamefont{Aichhorn}},
  \bibinfo{author}{\bibfnamefont{L.}~\bibnamefont{Vaugier}}, \bibnamefont{and}
  \bibinfo{author}{\bibfnamefont{S.}~\bibnamefont{Biermann}},
  \bibinfo{journal}{Phys. Rev. Lett.} \textbf{\bibinfo{volume}{107}},
  \bibinfo{pages}{266404} (\bibinfo{year}{2011}).

\bibitem[{\citenamefont{Crawford et~al.}(1994)\citenamefont{Crawford,
  Subramanian, Harlow, Fernandez-Baca, Wang, and Johnston}}]{Crawford1994}
\bibinfo{author}{\bibfnamefont{M.}~\bibnamefont{Crawford}},
  \bibinfo{author}{\bibfnamefont{M.}~\bibnamefont{Subramanian}},
  \bibinfo{author}{\bibfnamefont{R.}~\bibnamefont{Harlow}},
  \bibinfo{author}{\bibfnamefont{J.}~\bibnamefont{Fernandez-Baca}},
  \bibinfo{author}{\bibfnamefont{Z.}~\bibnamefont{Wang}}, \bibnamefont{and}
  \bibinfo{author}{\bibfnamefont{D.}~\bibnamefont{Johnston}},
  \bibinfo{journal}{Phys. Rev. B} \textbf{\bibinfo{volume}{49}},
  \bibinfo{pages}{9198} (\bibinfo{year}{1994}).

\bibitem[{\citenamefont{Ahn et~al.}(2015)\citenamefont{Ahn, Lee, and
  Kune\v{s}}}]{Ahn2015}
\bibinfo{author}{\bibfnamefont{K.-H.} \bibnamefont{Ahn}},
  \bibinfo{author}{\bibfnamefont{K.-W.} \bibnamefont{Lee}}, \bibnamefont{and}
  \bibinfo{author}{\bibfnamefont{J.}~\bibnamefont{Kune\v{s}}},
  \bibinfo{journal}{J. Phys.: Condens. Matter} \textbf{\bibinfo{volume}{27}},
  \bibinfo{pages}{085602} (\bibinfo{year}{2015}).

\bibitem[{\citenamefont{Subramanian et~al.}(1994)\citenamefont{Subramanian,
  Crawford, Harlow, Ami, Fernandez-Baca, Wang, and Johnston}}]{Rh214_1994}
\bibinfo{author}{\bibfnamefont{M.~A.} \bibnamefont{Subramanian}},
  \bibinfo{author}{\bibfnamefont{M.~K.} \bibnamefont{Crawford}},
  \bibinfo{author}{\bibfnamefont{R.~L.} \bibnamefont{Harlow}},
  \bibinfo{author}{\bibfnamefont{T.}~\bibnamefont{Ami}},
  \bibinfo{author}{\bibfnamefont{J.~A.} \bibnamefont{Fernandez-Baca}},
  \bibinfo{author}{\bibfnamefont{Z.~R.} \bibnamefont{Wang}}, \bibnamefont{and}
  \bibinfo{author}{\bibfnamefont{D.~C.} \bibnamefont{Johnston}},
  \bibinfo{journal}{Physica C} \textbf{\bibinfo{volume}{235}},
  \bibinfo{pages}{743} (\bibinfo{year}{1994}).

\bibitem[{\citenamefont{Perry et~al.}(2006)\citenamefont{Perry, Baumberger,
  Balicas, Kikugawa, Ingle, Rost, Mercure, Maeno, Shen, and
  Mackenzie}}]{Rh_Perry2006}
\bibinfo{author}{\bibfnamefont{R.~S.} \bibnamefont{Perry}},
  \bibinfo{author}{\bibfnamefont{F.}~\bibnamefont{Baumberger}},
  \bibinfo{author}{\bibfnamefont{L.}~\bibnamefont{Balicas}},
  \bibinfo{author}{\bibfnamefont{N.}~\bibnamefont{Kikugawa}},
  \bibinfo{author}{\bibfnamefont{N.~J.} \bibnamefont{Ingle}},
  \bibinfo{author}{\bibfnamefont{A.}~\bibnamefont{Rost}},
  \bibinfo{author}{\bibfnamefont{J.~F.} \bibnamefont{Mercure}},
  \bibinfo{author}{\bibfnamefont{Y.}~\bibnamefont{Maeno}},
  \bibinfo{author}{\bibfnamefont{Z.~X.} \bibnamefont{Shen}}, \bibnamefont{and}
  \bibinfo{author}{\bibfnamefont{A.~P.} \bibnamefont{Mackenzie}},
  \bibinfo{journal}{New J. Phys.} \textbf{\bibinfo{volume}{8}},
  \bibinfo{pages}{175} (\bibinfo{year}{2006}).

\bibitem[{\citenamefont{Moon et~al.}(2006)\citenamefont{Moon, Kim, Kim, Lee,
  Kim, Park, Kim, Oh, Nakatsuji, Maeno et~al.}}]{Rh_Moon2006}
\bibinfo{author}{\bibfnamefont{S.~J.} \bibnamefont{Moon}},
  \bibinfo{author}{\bibfnamefont{M.~W.} \bibnamefont{Kim}},
  \bibinfo{author}{\bibfnamefont{K.~W.} \bibnamefont{Kim}},
  \bibinfo{author}{\bibfnamefont{Y.~S.} \bibnamefont{Lee}},
  \bibinfo{author}{\bibfnamefont{J.-Y.} \bibnamefont{Kim}},
  \bibinfo{author}{\bibfnamefont{J.-H.} \bibnamefont{Park}},
  \bibinfo{author}{\bibfnamefont{B.~J.} \bibnamefont{Kim}},
  \bibinfo{author}{\bibfnamefont{S.-J.} \bibnamefont{Oh}},
  \bibinfo{author}{\bibfnamefont{S.}~\bibnamefont{Nakatsuji}},
  \bibinfo{author}{\bibfnamefont{Y.}~\bibnamefont{Maeno}},
  \bibnamefont{et~al.}, \bibinfo{journal}{Phys. Rev. B}
  \textbf{\bibinfo{volume}{74}}, \bibinfo{pages}{113104}
  (\bibinfo{year}{2006}).

\bibitem[{\citenamefont{Itoh et~al.}(1995)\citenamefont{Itoh, Shimura, Inaguma,
  and Morii}}]{Itoh1995}
\bibinfo{author}{\bibfnamefont{M.}~\bibnamefont{Itoh}},
  \bibinfo{author}{\bibfnamefont{T.}~\bibnamefont{Shimura}},
  \bibinfo{author}{\bibfnamefont{Y.}~\bibnamefont{Inaguma}}, \bibnamefont{and}
  \bibinfo{author}{\bibfnamefont{Y.~J.} \bibnamefont{Morii}},
  \bibinfo{journal}{Solid State Chem.} \textbf{\bibinfo{volume}{118}},
  \bibinfo{pages}{20} (\bibinfo{year}{1995}).

\bibitem[{\citenamefont{Chikara et~al.}(2015)\citenamefont{Chikara, Haskel,
  Sim, Kim, Chen, Fabbris, Veiga, Souza-Neto, Terzic, Butrouna
  et~al.}}]{Rh_Chikara2015}
\bibinfo{author}{\bibfnamefont{S.}~\bibnamefont{Chikara}},
  \bibinfo{author}{\bibfnamefont{D.}~\bibnamefont{Haskel}},
  \bibinfo{author}{\bibfnamefont{J.-H.} \bibnamefont{Sim}},
  \bibinfo{author}{\bibfnamefont{H.-S.} \bibnamefont{Kim}},
  \bibinfo{author}{\bibfnamefont{C.-C.} \bibnamefont{Chen}},
  \bibinfo{author}{\bibfnamefont{G.}~\bibnamefont{Fabbris}},
  \bibinfo{author}{\bibfnamefont{L.~S.~I.} \bibnamefont{Veiga}},
  \bibinfo{author}{\bibfnamefont{N.~M.} \bibnamefont{Souza-Neto}},
  \bibinfo{author}{\bibfnamefont{J.}~\bibnamefont{Terzic}},
  \bibinfo{author}{\bibfnamefont{K.}~\bibnamefont{Butrouna}},
  \bibnamefont{et~al.}, \bibinfo{journal}{Phys. Rev. B}
  \textbf{\bibinfo{volume}{92}}, \bibinfo{pages}{081114(R)}
  (\bibinfo{year}{2015}).

\bibitem[{\citenamefont{Liu et~al.}(2016)\citenamefont{Liu, Kim, Kumari,
  Mahadevan, Kresse, Sarma, Chen, and Franchini}}]{liu2016}
\bibinfo{author}{\bibfnamefont{P.}~\bibnamefont{Liu}},
  \bibinfo{author}{\bibfnamefont{B.}~\bibnamefont{Kim}},
  \bibinfo{author}{\bibfnamefont{P.}~\bibnamefont{Kumari}},
  \bibinfo{author}{\bibfnamefont{P.}~\bibnamefont{Mahadevan}},
  \bibinfo{author}{\bibfnamefont{G.}~\bibnamefont{Kresse}},
  \bibinfo{author}{\bibfnamefont{D.}~\bibnamefont{Sarma}},
  \bibinfo{author}{\bibfnamefont{X.-Q.} \bibnamefont{Chen}}, \bibnamefont{and}
  \bibinfo{author}{\bibfnamefont{C.}~\bibnamefont{Franchini}}
  (\bibinfo{year}{2016}), \bibinfo{note}{unpublished}.

\end{thebibliography}

\end{document}